\documentclass[a4paper,onecolumn]{agn6}
\usepackage{natbib}
\usepackage{graphicx}
\usepackage[a4paper]{hyperref}
\idline{}{1}
\newcommand{\ltsima}{$\; \buildrel < \over \sim \;$}
\newcommand{\simlt}{\lower.5ex\hbox{\ltsima}}
\newcommand{\gtsima}{$\; \buildrel > \over \sim \;$}
\newcommand{\simgt}{\lower.5ex\hbox{\gtsima}}

\newcommand{\cgs}{ ${\rm erg~cm}^{-2}~{\rm s}^{-1}$} 
\newcommand{\lum}{\rm erg s$^{-1}$}

\def\lesssim{\mathrel{\hbox{\rlap{\hbox{\lower4pt\hbox{$\sim$}}}\hbox{$<$}}}}
\def\gtrsim{\mathrel{\hbox{\rlap{\hbox{\lower4pt\hbox{$\sim$}}}\hbox{$>$}}}}

\def\arcsec{\hbox{$^{\prime\prime}$}}

\def\ab1450{$AB_{1450(1+z)}$}

\def\xray{\hbox{X-ray}}

\def\oiii{\hbox{[O\ {\sc iii}}]}
\def\lsun{\hbox{$L_\odot$}}
\def\loiii{$L_{[O\ III]}$}
\newcommand\phn{\phantom{0}}%
\def\chandra{{\it Chandra\/}}

\def\heao1{{\it HEAO-1\/}}

\def\rosat{{\it ROSAT\/}}

\def\xmm{{XMM-{\it Newton\/}}}

\begin{document}
\title{Type~II quasar candidates in the Sloan Digital Sky Survey: evidence for 
X-ray obscuration 
}

\author{C. Vignali \inst{1,2}, D.M. Alexander \inst{3} \and 
A. Comastri \inst{2}
}

\institute{Dipartimento di Astronomia, Universit\`a degli Studi di Bologna, 
Via Ranzani 1, 40127 Bologna, Italy; \email{cristian.vignali@bo.astro.it} 
              \and  
INAF--Osservatorio Astronomico di Bologna, Via Ranzani 1, 40127 Bologna, Italy
              \and 
Institute of Astronomy, Madingley Road, Cambridge, CB3~0HA, UK 
}

\abstract{
Recent \xray\ surveys with \chandra\ and \xmm\ have found several genuine 
Type~II quasars, the long sought-after population of high-luminosity 
(high-redshift) Seyfert~2 galaxies, and a 
sizable number of Type~II quasar candidates. 
However, at present it is hard to know whether the on-going \xray\ surveys 
are providing a reliable and nearly complete census of the Type~II quasar 
population. 
In order to address this open issue and shed light on the broad-band 
properties of Type~II quasars, we used the sample of 291 
high-ionization narrow emission-line AGN at $z$=0.3--0.83 selected from the 
Sloan Digital Sky Survey by \citet{zak03}. 
Using archival \xray\ information, we were able to place constraints on the 
\xray\ emission of 17 of these objects (three \xray\ detections and 14 
upper limits). 
Using the \oiii$\lambda$5007 line luminosities to predict the intrinsic 
\xray\ power of the AGN, we found that at least 47\% of the objects in the 
observed sample provides evidence for the presence of \xray\ absorption 
(with $N_{\rm H}$$\gtrsim$10$^{22}$~cm$^{-2}$), including the four 
highest luminosity sources with predicted unobscured luminosities of 
$\approx$~10$^{45}$~\lum. 
For the only source with moderate-quality \xmm\ spectral data, 
all of the pieces of the puzzle 
(2--10~keV luminosity of $\approx$~4$\times10^{44}$~\lum, 
\hbox{$N_{\rm H}$$\approx$~1--3$\times10^{22}$~cm$^{-2}$}, and 
unobscured magnitude $M_{\rm B}\approx-$26) assure that this is 
a genuine Type~II quasar. 

  }
   \authorrunning{C. Vignali et al.}
   \titlerunning{On the X-ray properties of SDSS Type~II quasar candidates}
   \maketitle

\section{Introduction}
Type~II quasars, the long sought-after luminous analogs of Seyfert~2 galaxies 
(i.e., sources with high-ionization, narrow optical emission-line spectra) 
predicted by Unification models of Active Galactic Nuclei (AGN), represent 
a key ingredient of many AGN synthesis models for the \xray\ background 
(XRB; e.g., \citealt{com01,gil01}). 
In the \xray\ band these objects are expected to be highly luminous 
\hbox{($\gtrsim$~a~few~$\times10^{44}$~\lum)} and absorbed by column densities 
\hbox{$\gtrsim10^{22}$~cm$^{-2}$}. 
Before the advent of 
\chandra\ and \xmm, 
Type~II quasars formed quite an elusive class of sources. 
It was not rare for AGN to be mis-classified as Type~II quasars due to 
either limited spectral coverage or low signal-to-noise ratio optical 
spectra (see \citealt{hal99}). 
\xray\ surveys conducted over the last few years have proven to be 
effective in finding several genuine Type~II quasars 
(i.e., matching both the optical and \xray\ definitions; e.g., 
\citealt{nor02,ste02,fio03,gan04,cac04,szo04}) and many more Type~II quasar 
candidates (e.g., \citealt{cra01,mai02,bar03}). 
Multiwavelength observations of these objects generally support their 
Type~II quasar status. However, some questions regarding Type~II quasars 
still remain without an answer despite the significant progress 
accomplished 
by \xray\ surveys over the last few years, mainly 
{\em what is the fraction of genuine Type~II quasars among the hard \xray\ 
selected population?} 
At present, less than $\approx$~20\% of the hard \xray, presumably absorbed 
sources, have optical counterparts matching the Type~II quasar definition 
(e.g., \citealt{bar03}). 
However, in many cases the faint optical counterparts of \xray\ selected 
Type~II quasar candidates hamper the spectroscopic identification, hence 
providing {\em further uncertainty on whether or not the census of these 
objects from \xray\ surveys is complete}. 
Thus, the selection of an optically bright sample can complement the 
information provided by \xray\ surveys and shed light on the properties 
of the Type~II quasar population as a whole.  

Motivated by these considerations, 
we have used the sample of 291 high-ionization narrow emission-line AGN 
at redshift \hbox{0.3--0.83} selected by 
\citet[ hereafter Z03]{zak03}\defcitealias{zak03}{Z03} 
from the Sloan Digital Sky Survey (SDSS) spectroscopic data. 
The sample includes (in almost similar fractions) both Seyfert~2 galaxies and 
their higher luminosity ``cousins'', Type~II quasars; although not complete, 
this catalog represents the only published optically selected 
sample of Type~II quasar candidates in this redshift range. 
Here we review the main \xray\ properties of the SDSS Type~II quasar 
candidates; a detailed analysis is presented by \citet{vig04}. 

$H_{0}$=70~km~s$^{-1}$~Mpc$^{-1}$, $\Omega_{\rm M}$=0.3, and 
$\Omega_{\Lambda}$=0.7 are used throughout the paper. 

\section{X-ray data}
The \citetalias{zak03} catalog of Type~II quasar candidates was cross 
correlated with archival \rosat\ (PSPC, HRI, and \rosat\ 
All Sky Survey, RASS), \chandra, and \xmm\ observations. 
Most of the data presented in the following have been obtained from \rosat\ 
observations, given the larger areal coverage of archival \rosat\ fields. 
We found reliable \xray\ information (a detection or upper limit) 
for 17 sources (listed in Table~\ref{tab1} along with the 
derived \xray\ parameters). 
One of these 17 sources was found in an archival \xmm\ observation, while 
none of the \citetalias{zak03} sources lies in archival 
\chandra\ observations. \\
\begin{table*}
\centering
\caption{Properties of the \citet{zak03} Type~II AGN with 
\xray\ data and constraints on the column densities.}
\begin{tabular}{|@{~}c@{~~}c@{~~}c@{~}c@{~}c@{~}c@{~}c@{}c@{~}|}
\hline
Src$^{a}$ & SDSS~J & $z$ & \loiii & $L_{\rm HX}$$^{b}$ & $F_{\rm SX}$$^{c}$ & $N_{\rm H,z}$$^{d}$ & Type~II$^{e}$ \\
ID & & & (\lsun) & (2--10~keV) & (0.5--2~keV) & Range & QSO \\ 
%
%
 & & & & & & & \\
\hline
{\phn}34   & 021047.01$-$100152.9 & 0.540 & {\phn}9.79 & 45.1       & $<2.00$ & $>$6.0$\times10^{22}$--2.0$\times10^{23}$   &  $\ast$  \\ 
{\phn}55   & 023359.93$+$004012.7 & 0.388 & {\phn}8.17 & 43.5       & $<0.43$ & $>$1.1$\times10^{22}$--9.6$\times10^{22}$   &          \\ 
{\phn}59   & 024309.79$+$000640.3 & 0.414 & {\phn}7.95 & 43.3       & $<0.92$ & $>$7.0$\times10^{21}$--4.9$\times10^{22}$   &          \\ 
{\phn}68   & 025558.00$-$005954.0 & 0.700 & {\phn}8.51 & 43.9       & $<3.54$ & $>$0--2.0$\times10^{22}$                    &          \\
{\phn}70   & 025951.28$+$002301.0 & 0.505 & {\phn}8.53 & 43.9       & $<2.87$ & $>$0--4.7$\times10^{22}$                    &          \\
     130   & 084234.94$+$362503.1 & 0.561 &      10.10 & 45.5       & $<2.53$ & $>$8.0$\times10^{22}$--2.3$\times10^{23}$   &  $\ast$  \\ 
\bf  148   & 090933.51$+$425346.5 & 0.670 & {\phn}8.92 & 44.3       & $39.5$  & 0                                        &          \\ 
     152   & 092014.11$+$453157.3 & 0.402 & {\phn}9.04 & 44.4       & $<2.75$ & $>$1.3$\times10^{22}$--1.0$\times10^{23}$   &  $\ast$  \\ 
     174   & 100854.43$+$461300.7 & 0.544 & {\phn}8.32 & 43.7       & $<3.42$ & $>$0--2.3$\times10^{22}$                    &          \\
     188   & 104505.39$+$561118.4 & 0.428 & {\phn}9.08 & 44.4       & $<3.51$ & $>$8.5$\times10^{21}$--9.5$\times10^{22}$   &  $\ast$  \\ 
\bf  204   & 122656.48$+$013124.3 & 0.732 & {\phn}9.66 & 45.0       & $2.63$  &    2.3$\times10^{22}$--1.8$\times10^{23}$   &  $\ast$  \\ 
     $^{f}$
           &                      &       &            & 44.6       & $3.90$  & 7.5$\times10^{21}$--2.0$\times10^{22}$   &  $\ast$  \\ 
     208   & 123453.10$+$640510.2 & 0.594 & {\phn}8.77 & 44.1       & $<4.74$ & $>$0--4.1$\times10^{22}$                    &          \\
     209   & 124736.07$+$023110.7 & 0.487 & {\phn}8.59 & 43.9       & $<3.61$ & $>$0--4.6$\times10^{22}$                    &          \\
     212   & 130740.56$-$021455.3 & 0.425 & {\phn}8.92 & 44.3       & $<8.37$ & $>$0--4.9$\times10^{22}$                    &          \\ 
{\bf 239}$^{g}$   
           & 150117.96$+$545518.3 & 0.338 & {\phn}9.06 & 44.4       &  $30.3$ &    0--3.0$\times10^{23}$                    &          \\
     256   & 164131.73$+$385840.9 & 0.596 & {\phn}9.92 & 45.3       & $<2.44$ & $>$6.0$\times10^{22}$--2.1$\times10^{23}$   &  $\ast$  \\ 
     258   & 165627.28$+$351401.7 & 0.679 & {\phn}8.57 & 43.9       & $<1.53$ & $>$0--6.2$\times10^{22}$                    &          \\
\hline
\end{tabular}
\begin{minipage}{135mm}
{\sc Note} --- The \xray\ detected objects are shown in boldface. \\
$^{a}$ From Table~1 of \citetalias{zak03}. 
$^{b}$ Logarithm of the predicted 2--10~keV luminosity in units of \lum; 
the relative uncertainty is $\pm{0.6}$ (i.e., the 1$\sigma$ 
scatter in the \citealt{mul94} correlation between the \oiii\ 
line flux and the \hbox{2--10~keV} flux found for Seyfert\ 2 galaxies). 
This correlation has been applied to our data and used to predict the 
expected \hbox{0.5--2~keV} flux for each object; see $\S$3 for details. 
$^{c}$ Galactic absorption-corrected flux (or 3$\sigma$ upper 
limit) in the observed \hbox{0.5--2~keV} band, in units of 10$^{-14}$~\cgs. 
$^{d}$ The range of column densities reported here (in cm$^{-2}$) 
has been derived on a source-by-source basis by comparing the predicted 
\hbox{0.5--2~keV} fluxes (using the \citealt{mul94} correlation) 
with the observed \hbox{0.5--2~keV} fluxes (or 3$\sigma$ upper limits). 
For the sources that have not been detected by \rosat, the column densities 
should be considered as lower limits (hence the use of ``$>$''). 
%
%
For the three \xray\ detected sources, the column density range is broadly 
consistent with the absorption derived from the hardness-ratio analysis using 
\rosat\ data. 
$^{e}$ $\ast$ indicates the likely absorbed sources with 
predicted 2--10~keV luminosities $\gtrsim3\times10^{44}$~\lum, 
i.e., within the Type~II quasar locus. 
%
%
$^{f}$ \xmm\ observation of source \#204. 
The observed \hbox{0.5--2~keV} flux and de-absorbed \hbox{2--10~keV} 
luminosity shown here have been derived directly from \xray\ spectral fitting 
of \xmm\ data. 
$^{g}$ Detected in the RASS. 
\end{minipage}
\label{tab1}
\end{table*}

\begin{figure}
\centering
\resizebox{\hsize}{!}{\rotatebox[]{0}{\includegraphics{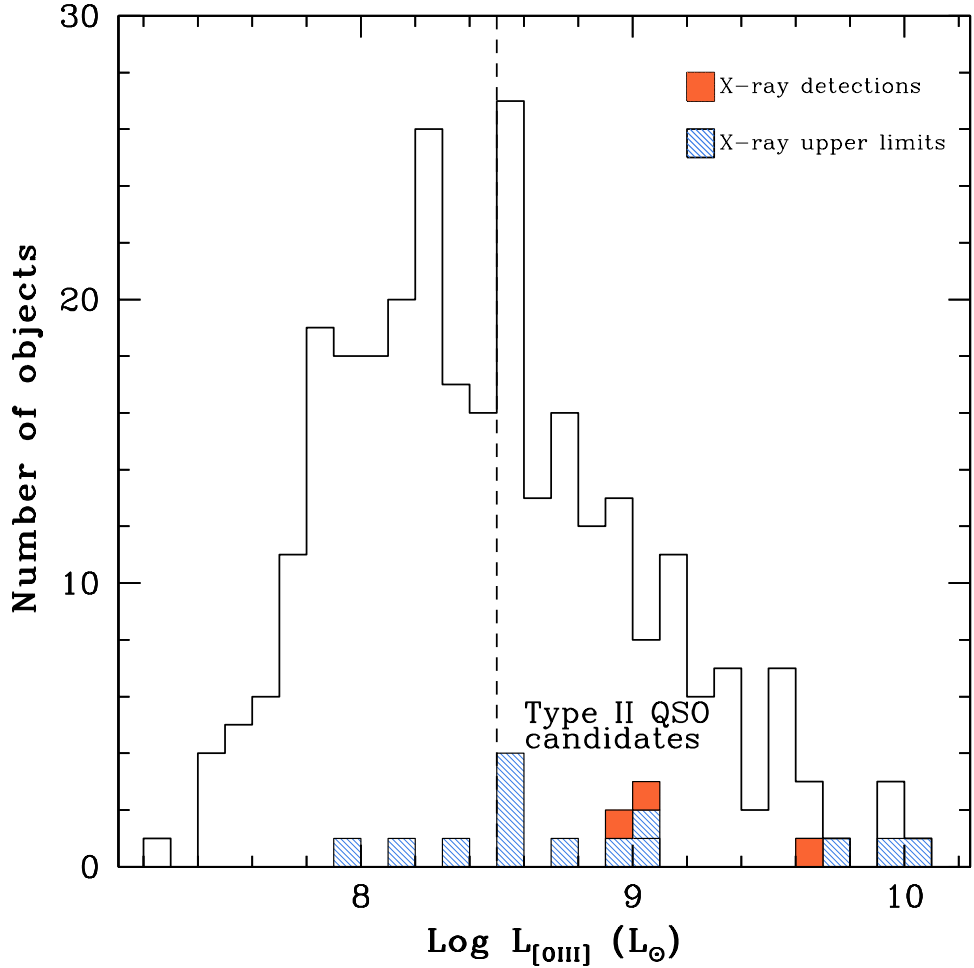}}}
\caption{Histogram of the \oiii\ line luminosities (uncorrected for absorption 
toward the narrow-line region) for all of the sources in 
the \citetalias{zak03} sample (thick solid line). 
The \loiii\ distributions for the sources with either \xray\ upper limits or 
detections are shown as filled histograms. 
The locus to the right of the vertical dashed line is populated by 
objects having \oiii\ line luminosities in the range 
$\approx$~3$\times10^{8}$ to $10^{10}$~$\lsun$, 
suggesting unobscured magnitudes of $-27<M_{\rm B}<-23$ 
(see \citetalias{zak03} for details).}
\label{fig1}
\end{figure}

Overall, adopting a detection threshold of 3.2$\sigma$ (3.7$\sigma$) 
for pointed (RASS) observations and a matching radius of 40\arcsec, 
three sources of the \citetalias{zak03} sample have significant 
($\simgt6\sigma$) \xray\ detections with \rosat\ 
(sources \#148, \#204, and \#239). 
Source \#204 was detected also by \xmm\ and has enough counts 
for moderate-quality spectral analysis (see $\S$4). 
The histogram of the \oiii\ line luminosities (\loiii) for all 
of the sources in the \citetalias{zak03} sample is shown in Fig.~\ref{fig1}. 
Note that 14 of the 17 sources with 
available \xray\ data (filled histograms) have \loiii\  
in the range \hbox{$\approx$~3$\times10^{8}-10^{10}$~$\lsun$}, 
suggesting unobscured magnitudes of $-27<M_{\rm B}<-23$ 
(see \citetalias{zak03} for details). 
We are also able to provide constraints on the \xray\ properties 
of four of the eight highest luminosity Type~II quasar candidates 
of the \citetalias{zak03} sample 
(predicted $L_{2-10~keV}$ luminosities of $\approx$~10$^{45}$~\lum; 
see Fig.~\ref{fig1} and Tab.~\ref{tab1}). 
The identification of such high-luminosity rare obscured quasars allows 
us to explore a different region of parameter space to that typically probed 
by \xray\ surveys. 
%

\section{The X-ray properties of Type~II quasar candidates}
To derive basic constraints on the \xray\ emission of the 17 
Type~II quasar candidates in our sample, 
we have used the correlation between the \oiii\ flux and the 
\hbox{2--10~keV} flux found for Seyfert~2 galaxies by 
\citet[ hereafter M94]{mul94}\defcitealias{mul94}{M94}adopting the 
parameterization reported in $\S$3.2 of \citet{col00}. 
Using this approach, for each source we have predicted the expected 
\hbox{2--10~keV} flux and errors (on the basis of the 1$\sigma$ scatter in the 
\citetalias{mul94} correlation). 
The [OIII] line luminosities used in this work are not 
corrected for the absorption toward the narrow-line region 
(see \citealt{mai98} for details). This correction would require the 
measurement of the Balmer decrement, but the redshift range of the 
\citetalias{zak03} sources and the bandpass of the SDSS spectra 
(3800--9200~\AA) do not provide spectral coverage of the 
H$\alpha$ line for all of the objects. 
We note that using the uncorrected [OIII] line flux 
leads to a conservative determination in whether or not a source 
is absorbed in the \xray\ band. 

The predicted \hbox{2--10~keV} fluxes (with their 1$\sigma$ uncertainties) 
have been converted into soft \hbox{(0.5--2~keV)} \xray\ fluxes assuming a 
power law with photon index $\Gamma$=2, which is a relatively good 
parameterization of the intrinsic \xray\ continua of both Type~I and 
Type~II AGN (e.g., \citealt{nan94,ris02}).
%
\begin{figure}
\centering
\resizebox{\hsize}{!}{\rotatebox[]{0}{\includegraphics{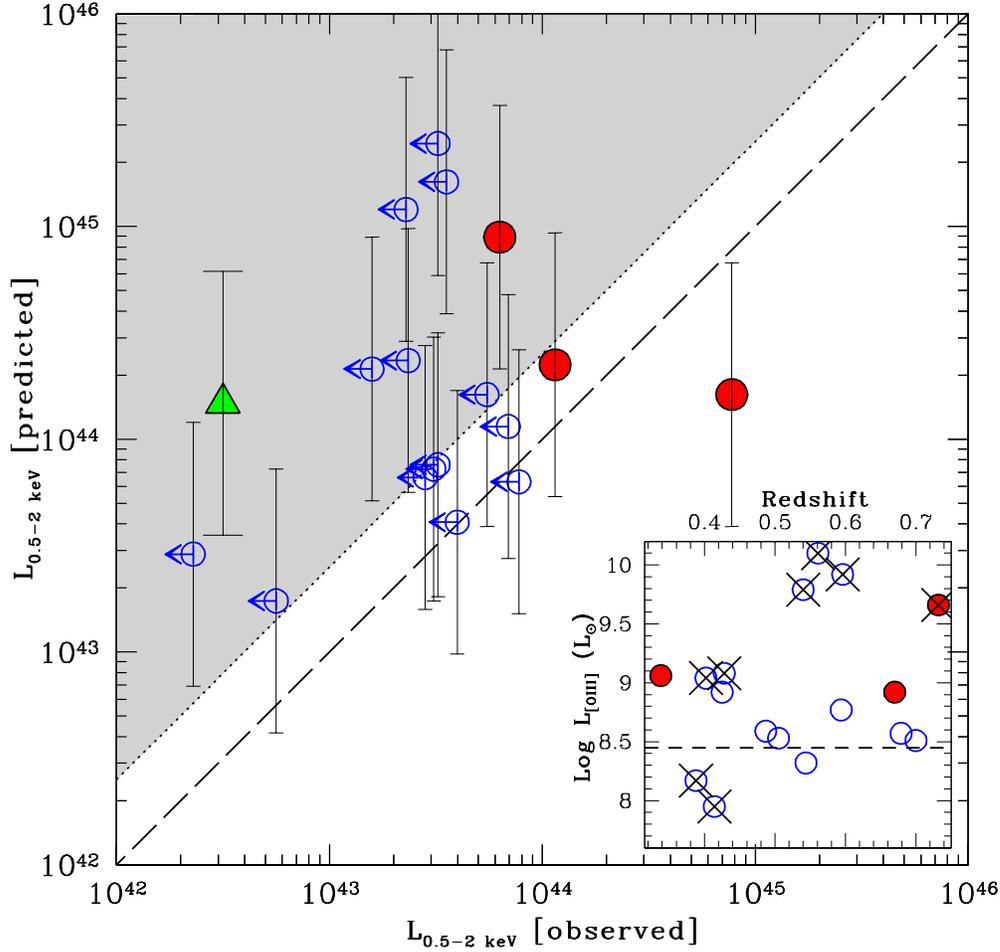}}}
\caption{Predicted vs. observed soft \xray\ luminosity for 
the \citetalias{zak03} objects with archival \xray\ observations. 
The filled circles indicate the three 
\xray\ detections (two from pointed observations and one from the RASS); 
the filled triangle shows the result from \xray\ stacking analysis 
of all \xray\ undetected objects (individually shown as open circles). 
Note the large \xray\ luminosity for the rightmost 
object (source \#148) of the plot; this is also the most radio-loud source 
of the entire \citetalias{zak03} catalog. 
The dashed diagonal line shows the 1:1 correlation, 
while the dotted line indicates the correlation expected for a source 
at $z$=0.5 if the column density is $10^{22}$~cm$^{-2}$. 
The grey region shows the locus of likely absorbed 
($N_{\rm H}>10^{22}$~cm$^{-2}$) sources. 
The column densities associated with our objects, given the 
method used for their derivation, should be considered conservative 
(see $\S$3 for details). 
In the inset, the \loiii\ vs. redshift is plotted for the objects with \xray\ 
observations. The crosses indicate the likely absorbed objects 
(see Table~\ref{tab1}). 
As in Fig.~\ref{fig1}, objects above the dashed line have 
\oiii\ line luminosities typical of unobscured quasars ($-27<M_{\rm B}<-23$).}
\label{fig2}
\end{figure}
%
These predicted soft \xray\ fluxes have then been compared to the 
observed soft \xray\ fluxes (or the 3$\sigma$ upper limits for 
the \xray\ non-detections; see Table~\ref{tab1}). 
This allows us to estimate on a source-by-source basis 
the amount of \xray\ absorption (listed as a range in 
Table~\ref{tab1} due to the scatter in the \citetalias{mul94} correlation) 
required to produce the predicted soft \xray\ flux. 
The principal, basic assumption used to derive the \xray\ column densities 
is that these Type~II quasar candidates have the same underlying average 
\xray\ continua of local AGN. 
Clearly, the uncertainties associated with the predicted soft \xray\ fluxes 
(and thus with the derived column densities) are large, mostly due to 
the scatter in the \citetalias{mul94} correlation and partly 
due to the assumed underlying \xray\ spectral slope. 
Intrinsically flatter \xray\ spectral slopes would 
produce lower column densities (by $\approx$~20\%, if $\Gamma$=1.6 instead of 
2.0 is adopted), while the presence of an additional soft \xray\ component 
(e.g., thermal emission from the host galaxy, scattering, etc.), often 
observed in hard \xray\ selected sources (e.g., \citealt{vig01}) would 
produce the opposite effect. 
However, {\sf within the assumptions adopted in this study, 
these column densities should be treated as lower limits}, 
since all but three of our sources are not detected in the soft \xray\ band. 

Using the approach described above, we have found that 
{\sf at least 47\% (8/17) of the sources with \xray\ information 
indicate the presence of \xray\ absorption} (see Fig.~\ref{fig2} 
and Table~\ref{tab1}), with column densities typically 
$\simgt10^{22}$~cm$^{-2}$ (dotted line in Fig.~\ref{fig2} for a source 
at $z$=0.5). 
The predicted \hbox{2--10~keV} luminosities of six of the eight 
likely absorbed sources place them in the quasar regime 
($L\gtrsim10^{44.4}$~\lum; see Table~\ref{tab1}). 
Furthermore, the four most luminous sources 
(\hbox{$L\approx10^{45}$~\lum}; see Table~\ref{tab1}) 
in the sample show evidence for obscuration at \xray\ energies. 
A column density of \hbox{$\approx$~1--3$\times10^{22}$~cm$^{-2}$} 
was obtained for source \#204 through direct \xray\ spectral 
analysis of \xmm\ data (see $\S$4). 

For the 14 \xray\ undetected sources we performed a stacking 
analysis, obtaining an $\approx$~4$\sigma$ detection 
(triangle in Fig.~\ref{fig2}). 
Although the statistical relevance of this detection is poor, the comparison 
of the average flux derived from the stacking analysis (weighted by the 
exposure time of each observation with respect to the total) with that 
expected on the basis of the \citetalias{mul94} correlation provides 
an average column density of \hbox{1.4--2.7$\times10^{23}$~cm$^{2}$}. 
Hopefully, tighter constraints will be obtained soon with the scheduled 
observations of some of the most luminous Type~II quasar candidates 
with \chandra. 

\section{The X-ray spectrum of a genuine Type~II quasar}
\begin{figure}
\centering
\resizebox{0.5\hsize}{!}{\rotatebox[]{0}{\includegraphics{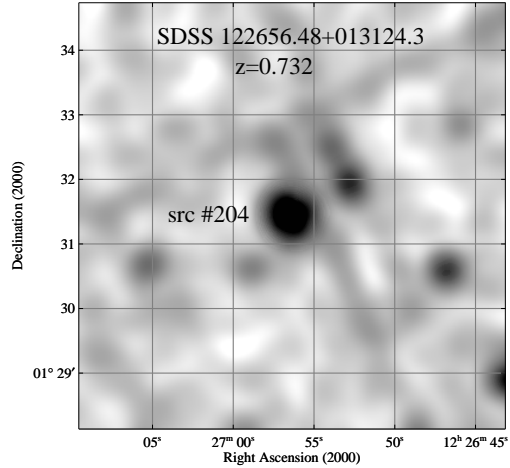}}}
\caption{Adaptively smoothed \xmm\ EPIC (pn$+$MOS1$+$MOS2) 
image centered on source \#204. 
The box size is $\approx$~400\arcsec$\times$400\arcsec. 
North is up, and East is to the left.}
\label{fig3}
\end{figure}

Source \#204 ($z$=0.732) has been serendipitously observed 
and detected by \xmm\ (see Fig.~\ref{fig3} 
for the combined pn$+$MOS1$+$MOS2 image), 
with a total number of \hbox{$\approx510$~counts} in the 0.5-10~keV band. 
%
%
A simple power-law model provides a poor fit to the \xmm\ data 
\hbox{($\chi^{2}$/d.o.f.=61.2/32)}; the derived flat photon index 
\hbox{($\Gamma$=0.74$^{+0.12}_{-0.13}$)} suggests the presence of absorption. 
%
\begin{figure}
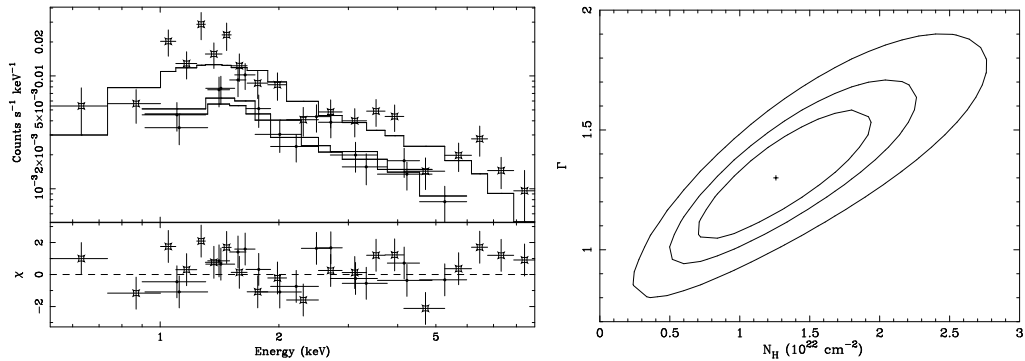

\centering
%
\includegraphics[angle=-90,width=0.52\textwidth]{vignali.fig4a.ps}\hfill
\includegraphics[angle=-90,width=0.46\textwidth]{vignali.fig4b.ps}
\caption{\xmm\ data of source \#204. Left panel: pn (larger open symbols) and 
MOS1$+$MOS2 (small filled circles) spectral data (top panel) and 
data/model residuals (bottom panel, in units of $\sigma$). 
Right panel: 68, 90, and 99\% confidence contours for the 
photon index vs. intrinsic column density.}
\label{fig4}
\end{figure}
%
If absorption at the source redshift is added to the power-law model, the fit 
improves significantly ($\Delta\chi^{2}$=16); 
the best-fitting spectrum (shown in Fig.~\ref{fig4}, left panel) 
is parameterized by a photon index \hbox{$\Gamma$=1.30$^{+0.16}_{-0.24}$} and 
a column density 
\hbox{$N_{\rm H}$=1.26$^{+0.75}_{-0.51}\times10^{22}$~cm$^{-2}$}. 
The 68, 90, and 99\% confidence contours for the photon index vs. column 
density are shown in Fig.~\ref{fig4} (right panel). 
If the photon index is forced to be \hbox{$\Gamma$=1.8--2.0}, 
more representative of the typical AGN intrinsic \xray\ emission, 
$N_{\rm H}$ becomes \hbox{2.2--3.2$\times10^{22}$~cm$^{-2}$}, 
more consistent with the lower column density boundary 
derived from \rosat\ data (see Table~\ref{tab1}). 
We note, however, that variability in the column density might have occurred 
between the \rosat\ and \xmm\ observations 
(over a time-scale of $\approx$~10~yr).
{\sf The intrinsic \hbox{2--10~keV} luminosity \hbox{(10$^{44.6}$~\lum)}, 
the column density \hbox{($\approx$~1--3$\times10^{22}$~cm$^{-2}$)}, 
the narrow-line optical spectrum (\citetalias{zak03}), and the 
predicted unobscured absolute $B$-band magnitude ($\approx-26$) 
are fully consistent with source \#204 being a genuine Type~II quasar. 
}
\section{Summary}
We have derived basic \xray\ constraints for 17 Type~II (high-ionization, 
narrow emission-line) quasar candidates at $z$=0.3--0.83 selected from the 
SDSS. 
\begin{description}
\item[(a)] 
At least 47\% (8/17) of the sources in the present sample indicate the 
presence of absorption, with column densities $\gtrsim10^{22}$~cm$^{-2}$. 
In particular, there are indications that the four highest luminosity 
objects \hbox{($L_{2-10~keV}$$\approx$~10$^{45}$~\lum)} are absorbed. 
The study of such objects is important since it allows the exploration 
of a different part of parameter space 
to that typically probed by \xray\ surveys. 
\item[(b)] 
X-ray spectral fitting of moderate-quality \xmm\ data for source \#204 
provides direct evidence of intrinsic absorption, with a column density of 
\hbox{$\approx$~1--3$\times10^{22}$~cm$^{-2}$}. This result, coupled with 
the intrinsic \hbox{2--10~keV} luminosity of 
\hbox{$\approx$~10$^{44.6}$~\lum}, the presence of luminous 
high-ionization and narrow optical lines, and the 
unobscured magnitude $M_{\rm B}\approx-$26 
makes this source a genuine 
Type~II quasar at optical and \xray\ wavelengths. 
\end{description}
Our results complement those obtained recently by 
\citet{zak04} on the basis of lower significance RASS data. 
In particular, \citet{zak04} noticed that the fraction of SDSS Type~I AGN 
at \hbox{$z$=0.3--0.8} with a RASS counterpart is much higher than that of 
SDSS Type~II AGN. This result is suggestive of suppression of soft 
\hbox{X-rays} by $N_{\rm H}$$\gtrsim10^{22}$~cm$^{-2}$ in the Type~II sources. 

Investigations of larger samples of SDSS Type~II quasars with 
\chandra\ and \xmm\ are clearly required to draw a more accurate and 
comprehensive picture of the physical properties 
of this enigmatic class of AGN.

\begin{acknowledgements}
CV and AC acknowledge partial support by the Italian Space agency 
(contract ASI I/R/057/02) and MIUR (COFIN grant 03-02-23). 
DMA is supported by the Royal Society. 
The authors would like to thank L. Angeretti for help in producing the plots. 
\end{acknowledgements}

\bibliographystyle{aa}

\begin{thebibliography}{}

\bibitem[Barger et al.(2003)]{bar03} Barger, A.J., et al. 
 2003, AJ, 126, 632
\bibitem[Caccianiga et al.(2004)]{cac04} Caccianiga, A., et al. 
 2004, \aap, 416, 901 
\bibitem[Collinge \& Brandt(2000)]{col00} Collinge, M.J., \& Brandt, W.N. 
 2000, \mnras, 317, L35
\bibitem[Comastri et al.(2001)]{com01} Comastri, A., Fiore, F., Vignali, C., 
 Matt, G., Perola, G.C., \& La Franca, F. 2001, \mnras, 327, 781 
\bibitem[Crawford et al.(2001)]{cra01} Crawford, C.S., Fabian, A.C., Gandhi, 
 P., Wilman, R.J., \& Johnstone R.M. 
 2001, \mnras, 324, 427 
\bibitem[Fiore et al.(2003)]{fio03} Fiore, F., et al. 
 2003, \aap, 409, 79
\bibitem[Gandhi et al.(2004)]{gan04} Gandhi, P., Crawford, C.S., Fabian, 
 A.C., \& Johnstone, R.M. 
 2004, \mnras, 348, 529 
\bibitem[Gilli, Salvati \& Hasinger(2001)]{gil01} Gilli, R., Salvati, M., 
 \& Hasinger, G. 
 2001, \aap, 366, 407
\bibitem[Halpern, Turner \& George(1999)]{hal99} Halpern, J., Turner, T.J., 
 \& George, I.M. 
 1999, \mnras, 307, L47
\bibitem[Mainieri et al.(2002)]{mai02} Mainieri, V., 
Bergeron, J., Hasinger, G., Lehmann, I., Rosati, P., Schmidt, M., Szokoly, 
 G., \& Della Ceca, R. 2002, \aap, 393, 425 
\bibitem[Maiolino et al.(1998)]{mai98} Maiolino, R., Salvati, M., Bassani, L., 
 Dadina, M., Della Ceca, R., Matt, G., Risaliti, G., \& Zamorani, G. 
 1998, \aap, 338, 781 
\bibitem[Mulchaey et al.(1994)]{mul94} Mulchaey, J.S., Koratkar, A., 
 Ward, M.J., Wilson, A.S., Whittle, M., Antonucci, R.R.J., Kinney, A.L., 
 \& Hurt, T. 1994, 
 \apj, 436, 586 (M94)
\bibitem[Nandra \& Pounds(1994)]{nan94} Nandra, K., \& Pounds, K.A. 
 1994, \mnras, 268, 405
\bibitem[Norman et al.(2002)]{nor02} Norman, C., et al. 
 2002, \apj, 571, 218
\bibitem[Risaliti(2002)]{ris02} Risaliti, G. 
 2002, \aap, 386, 379
\bibitem[Stern et al.(2002)]{ste02} Stern, D., et al. 
 2002, \apj, 568, 71
\bibitem[Szokoly et al.(2004)]{szo04} Szokoly, G.P., et al. 
 2004, \apjs, in press (astro-ph/0312324)
\bibitem[Vignali et al.(2001)]{vig01} Vignali, C., Comastri, A., Fiore, F., 
 \& La Franca, F. 2001, \aap, 370, 900
\bibitem[Vignali, Alexander \& Comastri(2004)]{vig04} Vignali, C., 
 Alexander, D.M., \& Comastri, A. 2004, \mnras, in press (astro-ph/0407293)
\bibitem[Zakamska et al.(2003)]{zak03} Zakamska, N.L., et al. 
 2003, \apj, 126, 2125 (Z03)
\bibitem[Zakamska et al.(2004)]{zak04} Zakamska, N.L., Strauss, M.A., 
 Heckman, T.M., Ivezic, Z., \& Krolik, J.H. 
 2004, \aj, 128, 1002

\end{thebibliography}

\end{document}